\documentclass[epj]{webofc}
\usepackage[varg]{txfonts}   
\woctitle{Hot Planets and Cool Stars}
\begin{document}
\title{The APACHE Project}
%
%

\author{A. Sozzetti\inst{1}\fnsep\thanks{\email{sozzetti@oato.inaf.it}}\and
        A. Bernagozzi\inst{2} \and
        E. Bertolini\inst{2}\and
         P. Calcidese\inst{2}\and
          A. Carbognani \inst{2}\and
           D. Cenadelli \inst{2}\and
           J.M. Christille \inst{2,3}\and
           M. Damasso\inst{1,2,4}\and
            P. Giacobbe\inst{1,2,5}\and
             L. Lanteri\inst{1}\and
              M.G. Lattanzi\inst{1}\and
                R. Smart\inst{1} 
}

\institute{INAF - Osservatorio Astrofisico di Torino Via Osservatorio 20, I-10025 Pino Torinese, Italy\and
Astronomical Observatory of the Autonomous Region of the Aosta Valley, Loc. Lignan 39, I-11020 Nus (Aosta), Italy\and
Dept. of Physics, University of Perugia, Via A. Pascoli, 06123 Perugia, Italy  \and
Dept. of Physics and Astronomy, University of Padova, Vicolo dell'Osservatorio 5, I-35122 Padova, Italy\and
Dept. of Physics, University of Trieste, Via Tiepolo 11, I-34143 Trieste, Italy
          }

\abstract{%
  First, we summarize the four-year long efforts undertaken to build the final setup of the APACHE Project, a photometric 
  transit search for small-size planets orbiting bright, low-mass M dwarfs. Next, we describe the present status of the APACHE survey, 
  officially started in July 2012 at the site of the Astronomical Observatory of the Autonomous Region of the Aosta Valley, in the 
  Western Italian Alps. Finally, we briefly discuss the potentially far-reaching consequences of a multi-technique characterization 
  program of the (potentially planet-bearing) APACHE targets. 
}
\maketitle

\section{Introduction} 

M dwarf stars, with masses $M_\star\leq0.6$ M$_\odot$, make up the vast 
majority of the reservoir of nearby stars within $\sim 25-30$ pc. These stars have not traditionally 
been included in large numbers in the target lists of 
radial-velocity (RV) searches for planets for two main reasons: 1) their
intrinsic faintness, which prevented Doppler surveys in the optical from
achieving very high radial-velocity precision ($<5-10$ m/s) for large samples 
of M dwarfs (e.g., Eggenberger \& Udry 2010, and references therein), and 2) their being
considered as providers of very inhospitable environments for potentially
habitable planets (e.g., Tarter et al. 2007; Scalo et al. 2007, and references therein). 
These two paradigms are now shifting. There are several important reasons for such a change in
perspective, which can be summarized under two main themes. 

First, the observational evidence gathered by 
ultra-high-precision space-borne photometric surveys (e.g., Kepler) indicates that the frequency of low-mass planets, 
i.e. Neptunes and Super-Earths, is an increasing function of decreasing stellar mass (Howard et al. 2012; Dressing \& Charbonneau 2013).
This result has recently 
been strengthened by the findings of ground-based RV programs carried out with state-of-the-art facilities 
(e.g., HARPS): Super-Earths with $M_p\leq 10$ $M_\oplus$ within the Habitable Zone\footnote{In its standard definition, the Habitable 
Zone corresponds to the range of distances from a given star for which water could be found in liquid form on a planetary surface 
(Kasting et al. 1993)} (HZ) of low-mass stars appear ubiquitous (Bonfils et al. 2013). Given that the identification of a rocky, 
habitable planet is the essential prerequisite to its possible characterization as an actual life-bearing celestial object, it is thus 
clear why low-mass M dwarfs, seen for long as providers of inhospitable environments for life (Scalo et al. 2007, and references therein), 
are now being moved at the center of the stage in the exoplanets arena. Consensus is growing among the astronomers' community that the first 
habitable rocky planet will be discovered (and might have been discovered already!) around a red M dwarf in the backyard of our Solar System. 

Second, the sample of the nearest ($d< 25-30$ pc), relatively bright ($J<9-10$) M dwarfs is amenable to combined studies with a wide array 
of observational techniques, which can be exploited at the best of their potential providing the opportunity to characterize the architecture 
of planetary systems across orders of magnitude in mass and orbital separations in a way that's not readily achievable for Solar analogs. 
For example, the possibility to reach detection of short-period transiting rocky planets from the ground with modest-size telescopes 
($30-50$ cm class) is guaranteed by the small radii of M dwarfs, leading to deep transits ($\Delta mag\gtrsim 0.005$ mag) for the case of 
planets with $2\leq R_p\leq 4$ $R_\oplus$ (e.g., Charbonneau et al. 2009). In addition, as we have discussed above, the favorable mass 
ratios allow for detection of rocky, potentially habitable planets with the RV technique, thanks to the moderately large amplitudes of the 
RV signals (a few m/s). Analogously, at intermediate separations ($\approx 1-4$ AU) high-precision astrometry becomes sensitive 
to planets in the mass range between Neptune and Jupiter (e.g., Casertano et al. 2008). Finally, the favorable planet-star contrast ratios 
provided by the low intrinsic luminosity of M dwarfs allows for improved detectability thresholds of giant planets at wide separations ($>5-10$ AU) 
with direct imaging techniques (e.g., Bowler et al. 2012). For the same reason, atmospheric characterization of transiting low-mass Super-Earths can be achieved 
for this sample (e.g., Tessenyi et al. 2012). Systematic studies of planetary systems orbiting low-mass stars can thus crucially inform 
planet formation and evolution, structural and atmospheric models, particularly when seen in connection with the properties of the hosts 
(e.g., Ida \& Lin 2005; Mordasini et al. 2012a,b; Seager \& Deming 2010, and references therein). 

However, not all physical properties of low-mass stars are known precisely enough for the purpose of the 
detection and characterization of small-radius planets. Worse still, some of the characteristics intrinsic to 
late-type dwarfs can constitute a significant source of confusion in the interpretation in planet detection 
and characterization measurements across a range of techniques. First of all, there exist 
discrepancies between theory and observations in the determination of the sizes of M dwarfs, 
typically on the order of 10\%-15\% (Ribas 2006; Torres et al. 2010, and references therein). 
Second, there are at present difficulties in spectroscopically determining with a high degree of precision 
M dwarf metallicities, as M-dwarf spectra are dominated by chemically complex molecular features that make 
the identification of the continuum in an M dwarf spectrum challenging, rendering line-based metallicity indicators unreliable 
(Woolf et al. 2009; Rojas-Ayala et al. 2010). Third, our understanding of the age-rotation-activity connection for field M dwarfs with age 
greater than $t\sim 0.5$ Gyr (e.g., Jenkins et al. 2009; West \& Basri 2009) is still subject to rather large uncertainties 
due to the sparseness of the data. 
Fourth, as measurements of chromospheric activity indicators (H$\alpha$ line) have shown how 
the fraction of active M dwarfs increases as a function of spectral sub-type (e.g., Bochanski et al. 2005; 
West et al. 2011), activity-related phenomena such as stellar spots, plages, and flares become increasingly a matter 
of concern for planet detection and characterization programs targeting late-type stars. 

\begin{figure}[t]
\centering
$\begin{array}{cc}
\includegraphics[width=0.44\textwidth]{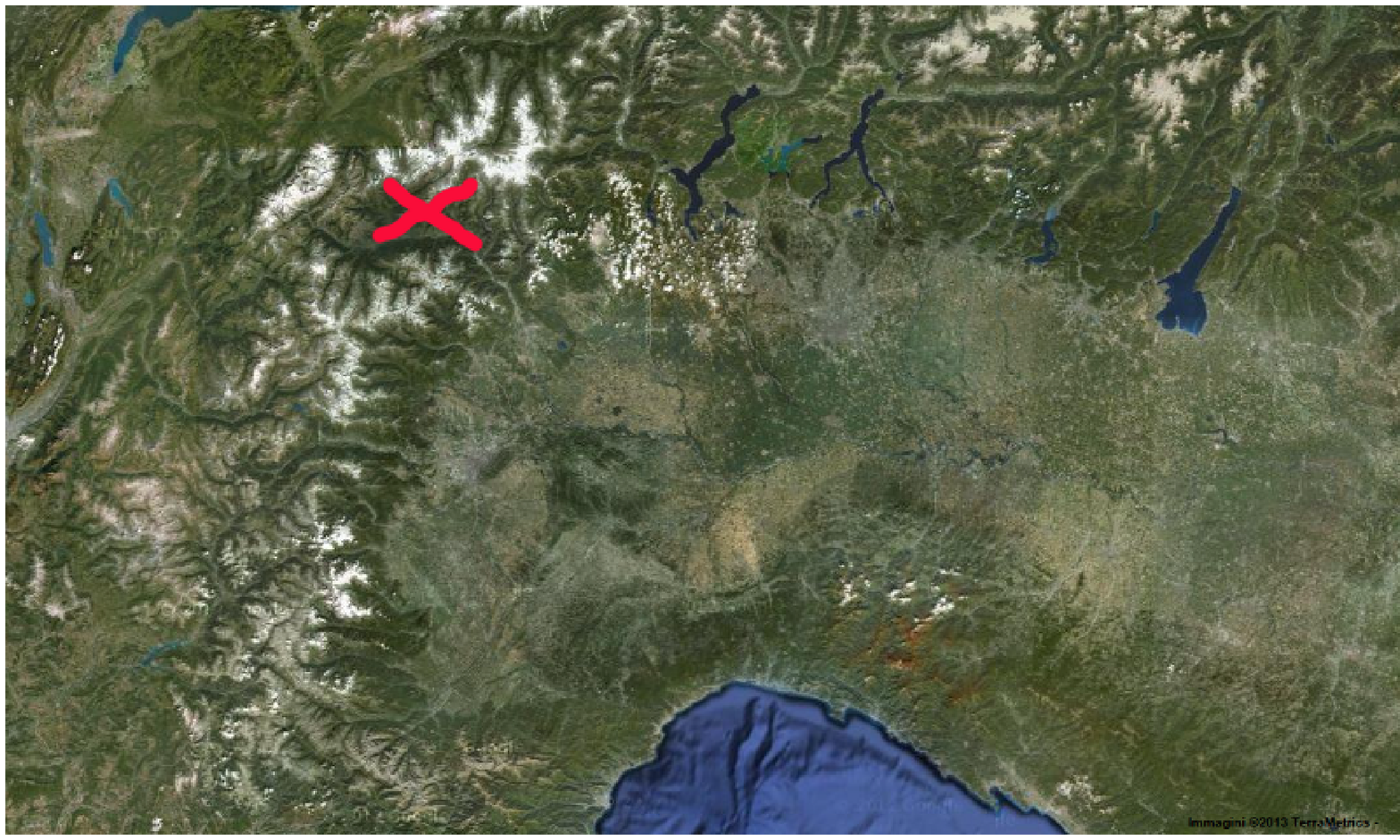} & 
\includegraphics[width=0.52\textwidth]{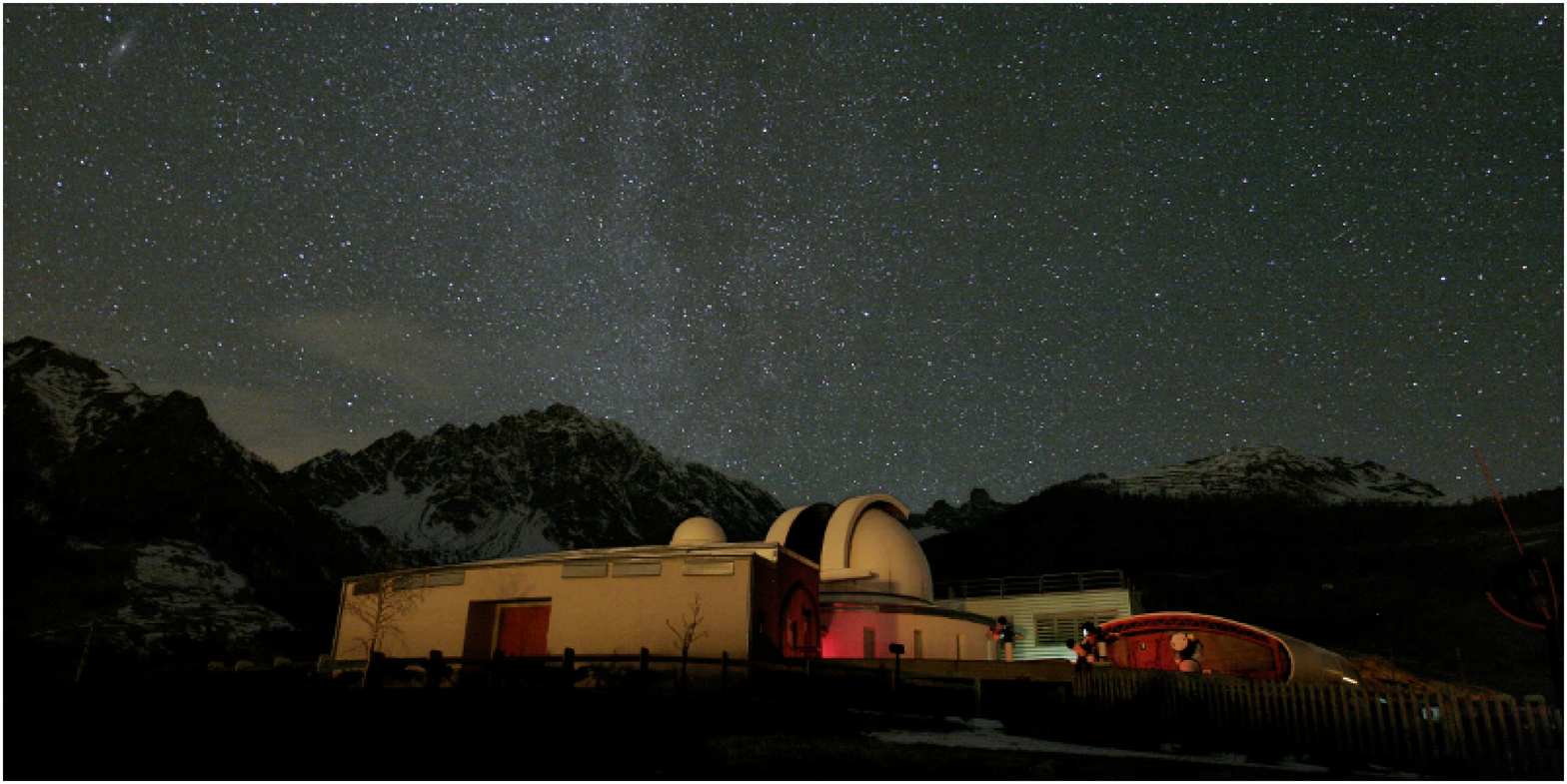} \\
\end{array} $
 \caption{Left: Map of northern Italy with the red cross marking the approximate location of the OAVdA in the Western Italian Alps. 
 Right: A night view of the Aosta Valley Observatory.}
\label{fig1}
\end{figure}

All the above considerations clearly underline how achieving the goal of the detection {\it and} characterization of low-mass, potentially habitable, 
rocky planets around low-mass stars requires the construction of a large (all-sky) sample of nearby, relatively bright M dwarfs with 
well-characterized properties. This will necessitate the combined use of time-series of spectroscopic, astrometric, and photometric 
data of high quality. In particular, the jitter levels will have to be quantified in detail for each target individually, 
as the jitter properties may vary from star to star within the same spectral class, as suggested by recent findings 
based on high-precision Kepler photometry (e.g., Ciardi et al. 2011) and high-resolution, high-S/N spectroscopy (e.g., Zechmeister et al. 2009). 
In the following, we summarize the preparations for and present first results from APACHE (A PAthway to the Characterization of Habitable Earths), 
a new ground-based photometric transiting search for small-size planets 
around thousands of bright, nearby early- to mid-M dwarfs. The APACHE project is a collaboration between INAF-Osservatorio Astrofisico di Torino (INAF-OATo) 
and the Astronomical Observatory of the Autonomous Region of the Aosta Valley (OAVdA). APACHE utilizes an array of five 40-cm telescopes 
at the OAVdA site, in the western Italian Alps (Fig.~\ref{fig1}, left and right panels). The survey, with official kick-off in July 2012, will last for five years. 

\section{The Site Characterization Study}

In this work (Damasso et al .2010), we carried out a detailed site characterization study at the OAVdA site, aimed at establishing its potential
to host the APACHE project. For the purpose of the site testing campaign, we gathered $R$-band photometric and seeing measurements utilizing
different instruments available at the site. We gauged site-dependent observing conditions such as night-sky brightness, photometric precision, 
and seeing properties. Public meteorological data were also used in order to help in the determination of the actual number of useful
observing nights per year. 

\begin{figure}[t]
\centering
$\begin{array}{cc}
\includegraphics[width=0.52\textwidth]{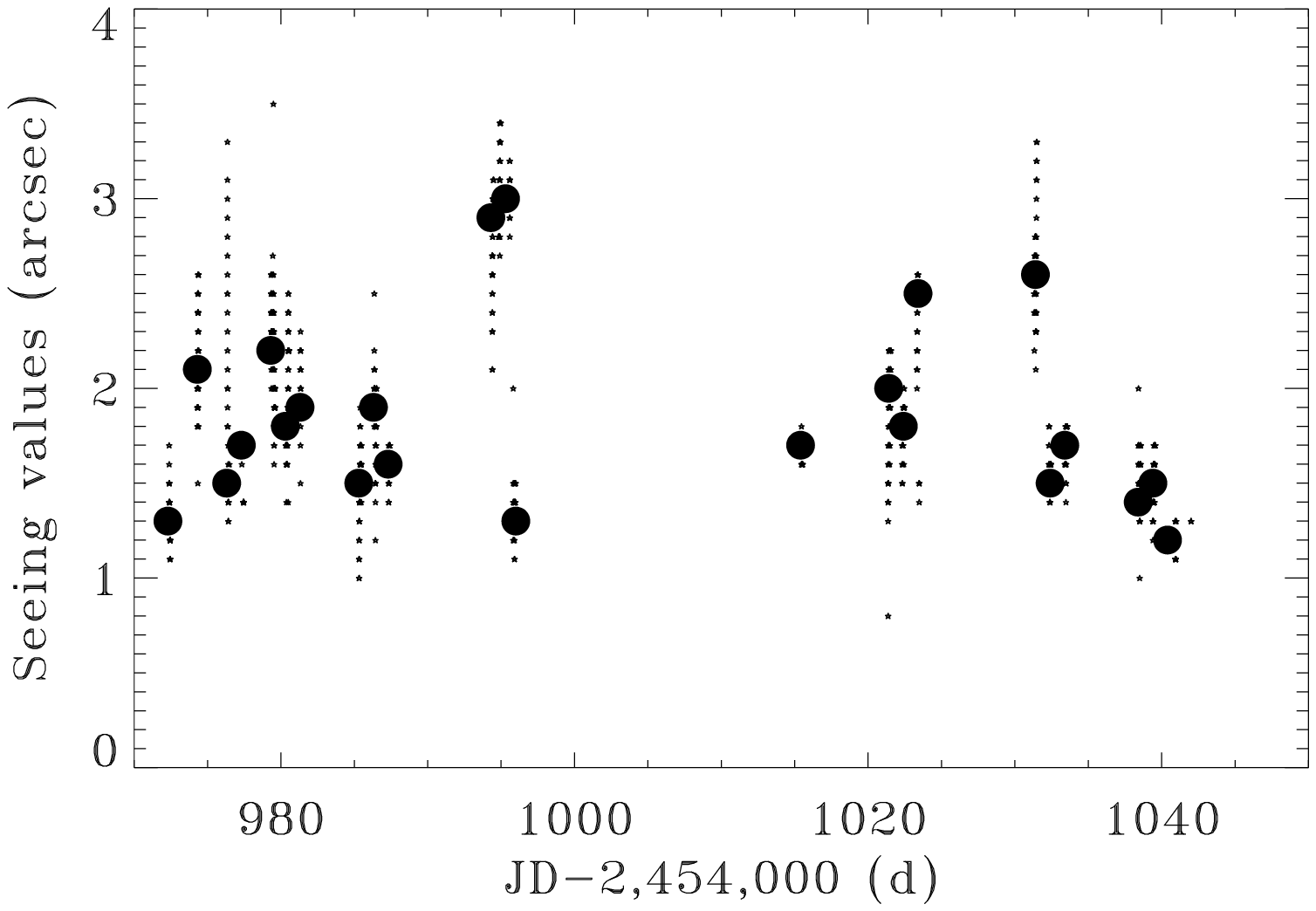} & 
\includegraphics[width=0.43\textwidth]{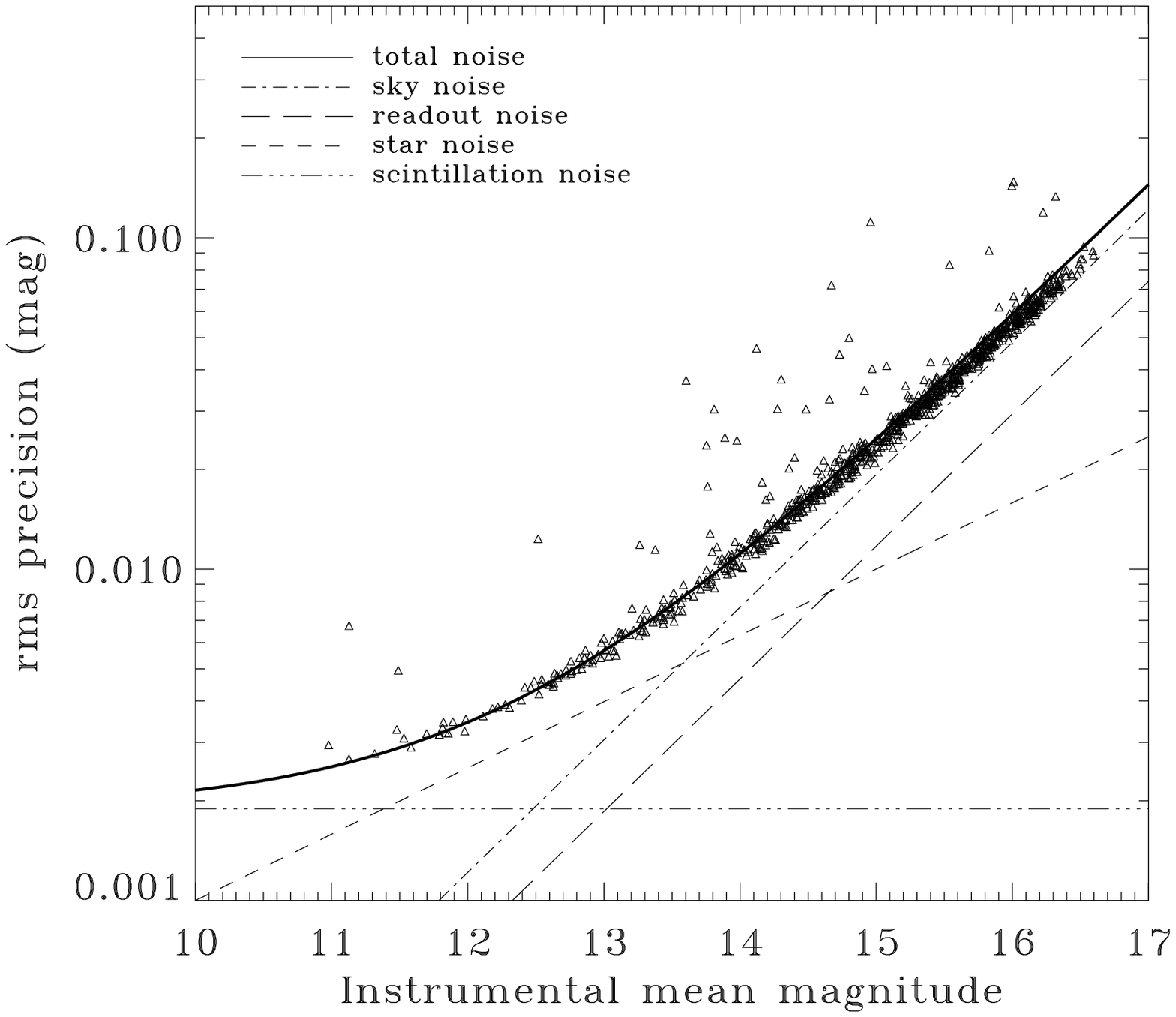} \\
\end{array} $
 \caption{Left: The distribution of seeing measurements over the whole period of
observation reported in Damasso et al. (2010). The small black stars show individual data points, each corresponding to an average seeing value over a 1 min
interval. The large filled circles indicate the median seeing value for each night. 
 Right: Photometric errors (rms) vs. instrumental mean magnitude in the
field of the planet-hosting star WASP-3 during a good observing night (Damasso et al. 2010). The various contributions to the expected photometric
noise, according to an analytic error budget model, are also shown. Pictures reprinted with permission from Damasso et al. (2010).}
\label{fig2}
\end{figure}

We measured $V$-band night-sky brightness and extinction levels not dissimilar from those registered at other 
good observing sites. The median seeing over the period of in situ observations (a few months) was found to be $\approx1.7^{\prime\prime}$ 
(Fig.~\ref{fig2}, left panel), and rather stable. However, given the limited duration of the observations, we
did not probe any possible seeing seasonal patterns or the details of its possible dependence on other meteorological
parameters, such as wind speed and direction. The fraction of fully clear nights per year amounts to 39\%, while the total of
useful nights increases to 57\%, assuming a (conservative) cloud cover of not more than 50\% of the night. Based on
the analysis of photometric data collected over the period of 2009 May-August for three stellar fields centered on the
transiting planet hosts WASP-3, HAT-P-7, and Gliese 436, we achieved seeing-independent best-case photometric
precision $\sim3$ mmag (rms) in several nights for bright ($R< 11$ mag) stars (see Fig.~\ref{fig2}, right panel for an example). 
A median performance $\sim6$ mmag during the observing period was obtained for stars with $R <13$ mag.

\section{The Pilot study}

The pilot study (Giacobbe et al. 2012) was focused on a year-long photometric monitoring campaign, 
using small-size instrumentation (25-80 cm class telescope systems), of a sample of $23$ bright ($J<12$ mag), 
cool M0-M6 dwarfs with good parallaxes from the TOrino Parallax Program (TOPP; Smart et al. 2010). 
The primary objectives we set out to achieve in this study were $a)$ to demonstrate
sensitivity to $<4$ R$_\oplus$ (i.e., smaller than radius of Neptune) transiting planets with periods of a few days around our
sample, through a two-fold approach that combines a characterization of the 
statistical noise properties of our photometry with the determination of transit detection 
probabilities via simulations, and $b)$ where possible, to better our knowledge 
of some astrophysical properties (e.g., activity, rotation, age) of our targets through a combination of 
spectroscopic and astrometric information and our differential photometric measurements. 

\begin{figure}[t]
\centering
$\begin{array}{cc}
\includegraphics[width=0.44\textwidth]{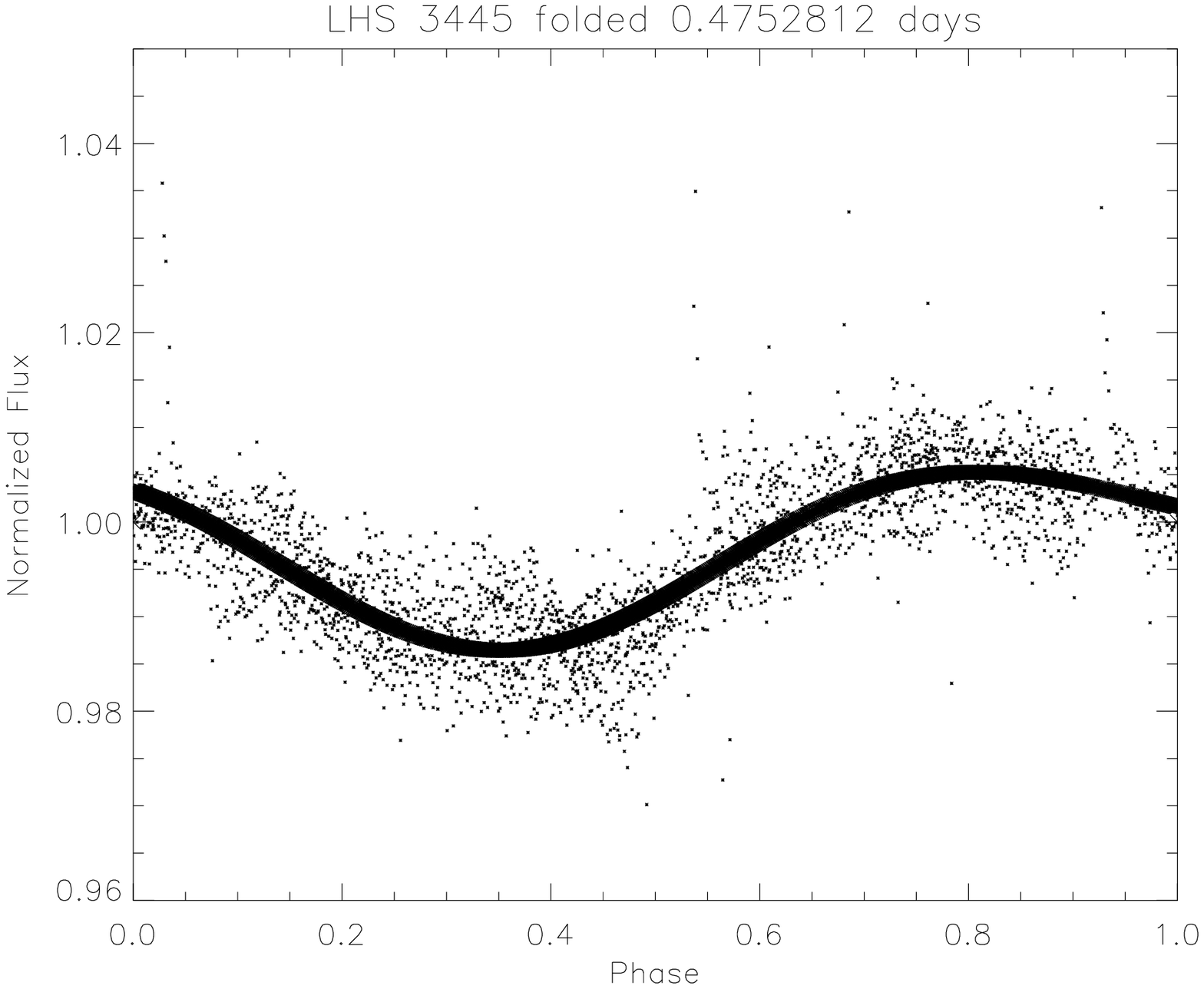} & 
\includegraphics[width=0.52\textwidth]{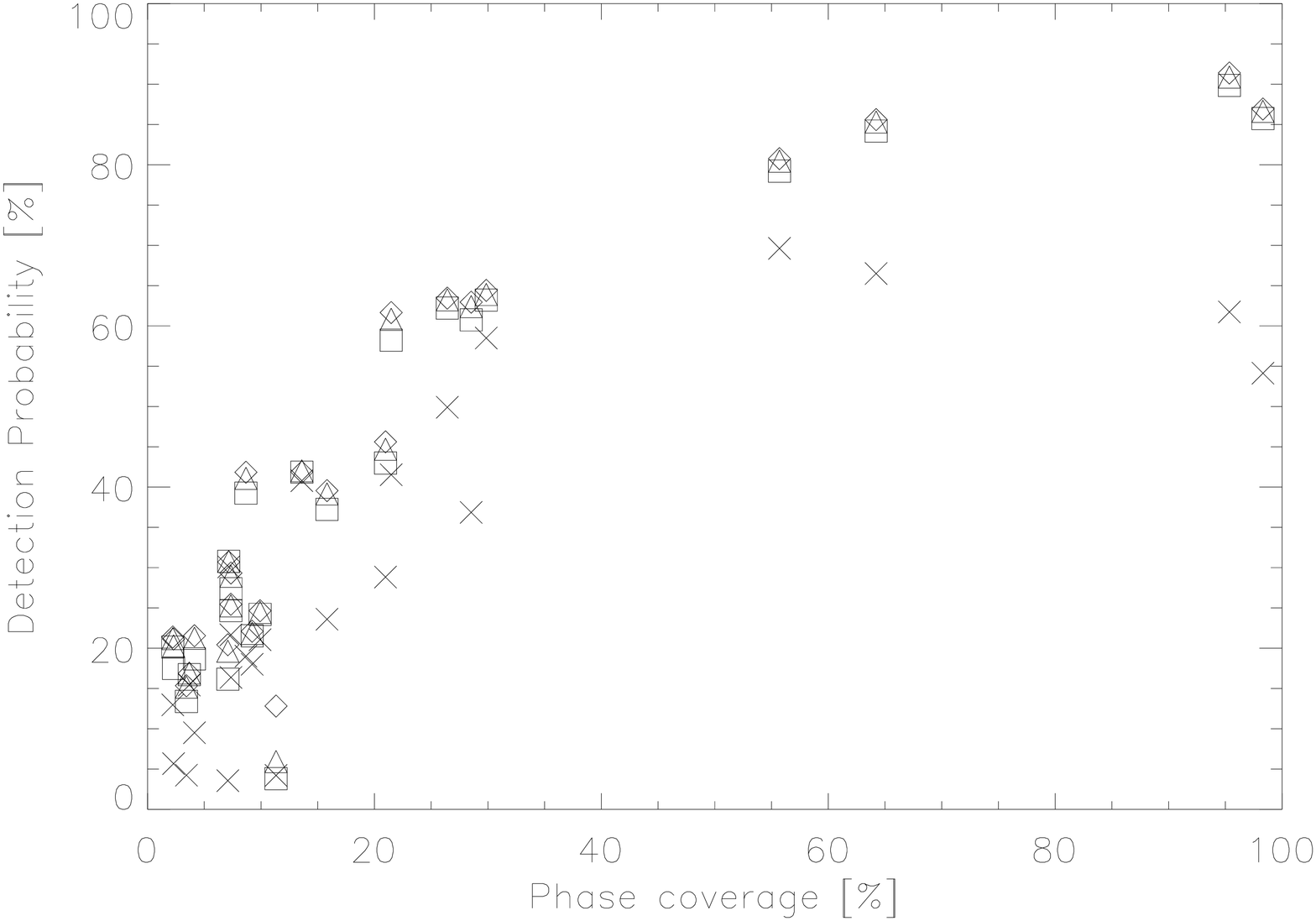} \\
\end{array} $
 \caption{Left: Photometric data for the dM star LHS 3445 folded according to its probable rotation period ($P=0.47$2 days). 
 Overplotted is a best-fit single-spot model, as discussed in Giacobbe et al. (2012). 
 Right: Average detection probability as a function of average phase coverage for $P < 1.0$ days. Pictures reprinted with 
permission from Giacobbe et al. (2012).}
\label{fig3}
\end{figure}

We achieved a typical nightly rms photometric precision of $\sim5$ mmag, with little or no dependence on the instrumentation used or on 
the details of the adopted methods for differential photometry, and gauged the degradation levels in the precision 
with respect to a pure white noise regime due to the presence of correlated (red) noise in our data. 
Based on a detailed stellar variability analysis, 
$a)$ we detected no transit-like events (an expected result given the sample size); $b)$ we determined $<1$ day 
photometric rotation periods for two of the M dwarfs in the sample, in agreement with the large projected rotational velocities 
inferred for both stars based on the analysis of archival spectra, and found consistency of the estimated inclinations of the 
stellar rotation axes with those derived using a simple spot model (see Fig.~\ref{fig3}, left panel). For two stars in the 
sample we recorded short-term, low-amplitude flaring events. Finally, based on simulations of transit signals of given period and 
amplitude injected in the actual (nightly reduced) photometric data for our sample, we derived 
a relationship between transit detection probability and phase coverage (see Fig.~\ref{fig3}, right panel). We found that, using the BLS 
search algorithm, even when phase coverage approached 100\%, there was a limit to the detection probability of $\approx 90\%$. 
Around program stars with phase coverage $>50\%$
we would have had $>80\%$ chances of detecting planets with $P<1$ day inducing fractional transit depths $>0.5\%$, 
corresponding to minimum detectable radii in the range $\sim1.0-2.2$ $R_\oplus$. 

\section{The APACHE Infrastructure}

\begin{figure}[t]
\centering
$\begin{array}{cc}
\includegraphics[width=0.44\textwidth]{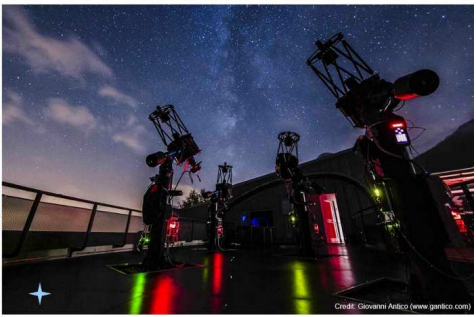} & 
\includegraphics[width=0.52\textwidth]{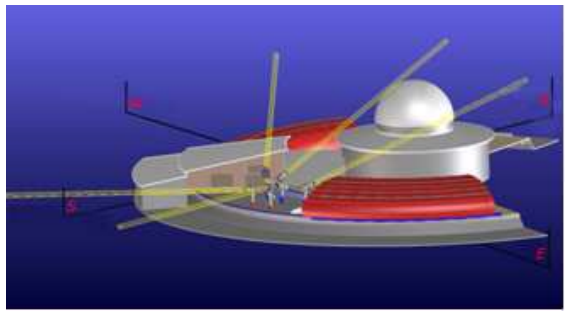} \\
\end{array} $
 \caption{Left: Close-up night view of the APACHE telescope array. 
 Right: A cartoon visualizing the exact arrangement of the APACHE telescopes within the OAVdA platform.}
\label{fig4}
\end{figure}

APACHE employs an array of five 40-cm class telescopes hosted on a single platform with electronically controlled roll-off enclosure 
(see Fig.~\ref{fig4}, left panel). The availability of a single movable enclosure
and of much of the associated support infrastructure makes the OAVdA a site essentially unique in Europe and will
afford a very substantial cost saving for the project. The arrangement of the array has been defined so as to maximize
sky coverage during campaign operations (see Fig.~\ref{fig4}, right panel). The telescope 
array is composed of five identical Carbon Truss 40-cm f/8.4 Ritchey-Chr\'etien telescopes, with a GM2000 10-MICRON mount and equipped with a 
FLI Proline PL1001E-2 CCD Camera and Johnson-Cousins R \& I filters. These systems guarantee state-of-the-art quality performance: 
negligible temperature gradients, QE~80\% in the whole wavelength range of interest. The open source observatory manager RTS2 
(Kubanek 2010) is the choice for the high-level software control of the five-telescope system, including dynamic scheduling of
the observations (for details, see Christille et al., this volume). 
Provisions have been made for the hardware of relevance for the APACHE operations control, data
acquisition, processing centre and backup, which is necessary to accommodate the high data rate: two HP
Z600 Workstation and one HP EliteBook 8740w Mobile Workstation as control PCs and workstations for data
analysis, one HP Storage Works X1600 12TB as Network Attached Storage (NAS), connected together via Intranet 1 GB. 

\section{Survey Design: Input Catalogue and Observing Strategy}

 \begin{figure}[t]
\centering
\includegraphics[width=0.85\textwidth]{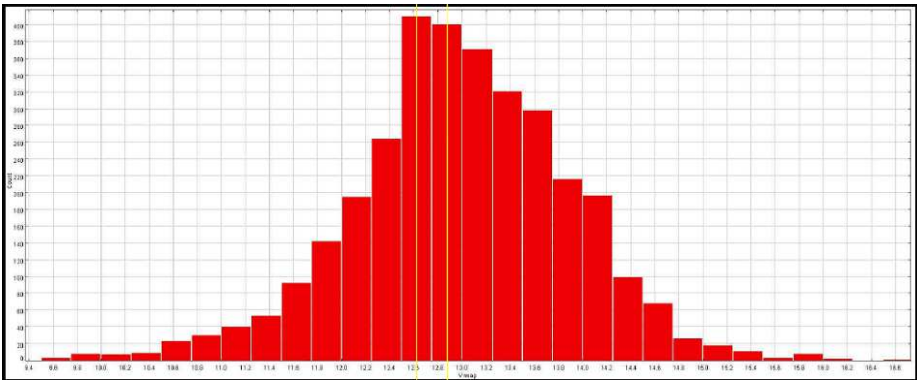}
 \caption{V mag distribution of the APACHE target sample.}
\label{fig5}
\end{figure}

The APACHE Input Catalogue (AIC) of M dwarfs is constructed starting from the all-sky sample of 8889 
bright ($J<10$) low-mass stars in L\'epine \& Gaidos (2011).
The final AIC is composed of $\sim3000$ targets selected on the basis of their visibility from the OAVdA site and 
the availability of a suitable number of potentially good comparison stars in the telescopes' field-of-view. 
Moreover, cross-correlation with approximately two dozen catalogues was performed to get further information about the targets, 
such as a precise determination of spectral class, their projected rotational velocity $V\sin i$, their level of chromospheric activity, 
X-ray emission, the presence of binary companions detected either spectroscopically or by direct imaging, or astrometry, or a 
combination of these techniques. Based on accurate parameterization of Gaia's scanning law, the number of the Gaia transits scheduled 
was associated to each target. Based on all available information, a prioritized list was defined for the purpose of scheduling 
with the APACHE telescopes. The final optical magnitude distribution of the sample peaks at $V\approx13$ mag, with a 
median of $V\sim14$ mag (Fig.~\ref{fig5}). The vast majority of targets is bright enough to allow for a host of additional 
characterization measurements, in case of positive detections of transiting planets. As a synergetic example, a carefully selected sub-sample of APACHE targets 
is currently the objective of high-precision RV monitoring with the HARPS-N spectrograph on the Telescopio Nazionale Galileo (TNG) as part of the 
long-term Global Architecture of Planetary Systems (GAPS) programme.

We built upon the experience matured with the pilot study to define and then refine the details of the observing strategy for the APACHE 
survey. Taking into account the use of synthetic transits, by assuming different numbers of consecutive exposures (from 1 to 4) and different 
temporal samplings (intervals of 20 through 50 minutes between two sets of consecutive pointings of the telescope on the same target), 
we evaluated the transit detection probabilities for stars with different average phase coverages 
in a given interval of possible orbital periods. 
The results of this analysis prompted us to adopt an initial observing strategy consisting of 3 exposures every $\sim20$ minutes, 
to be eventually revisited as the survey unfolds.

\section{The First Six Months of Operations}
 \begin{figure}[t]
\centering
$\begin{array}{cc}
\includegraphics[width=0.40\textwidth]{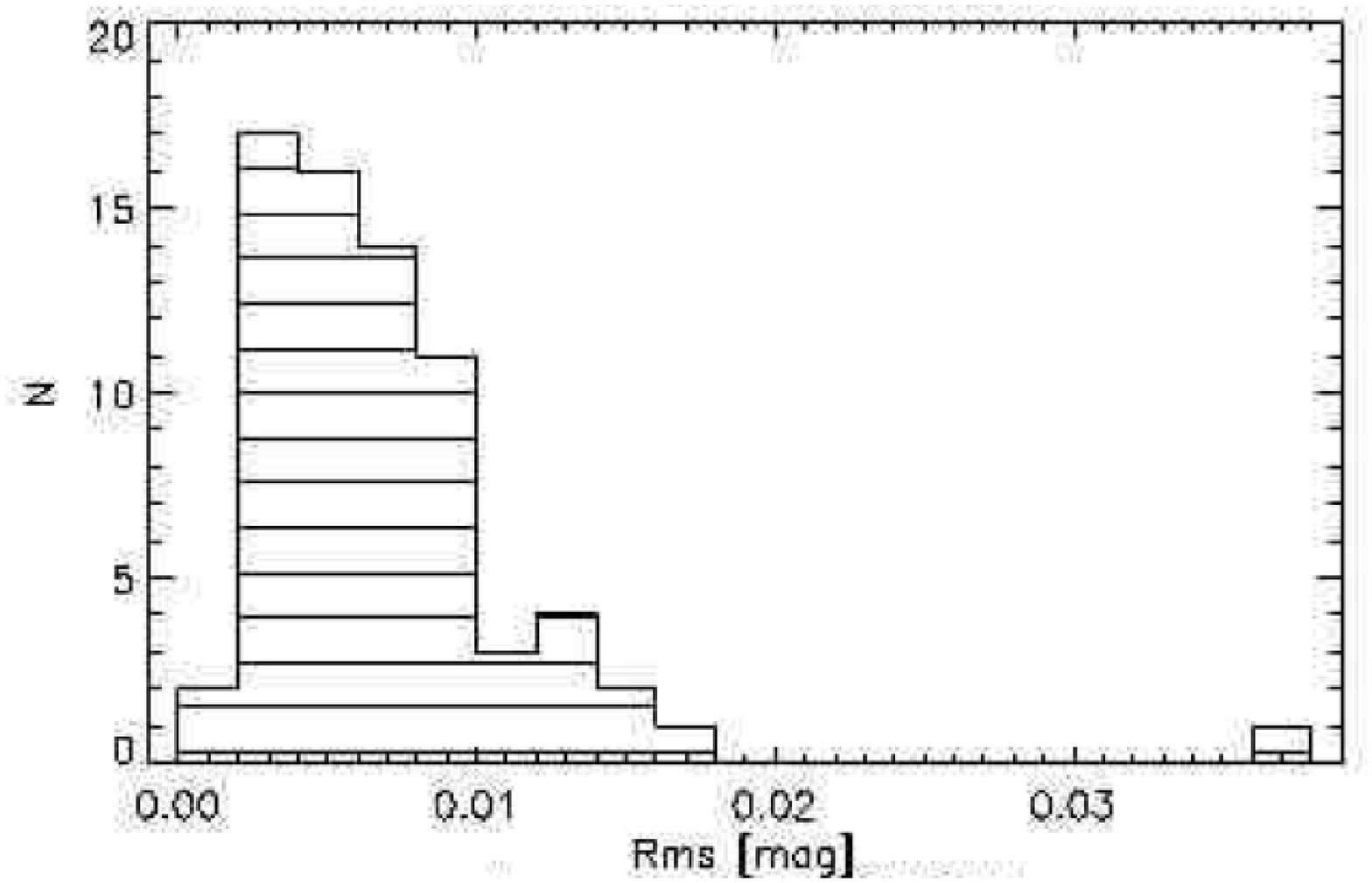} & 
\includegraphics[width=0.57\textwidth]{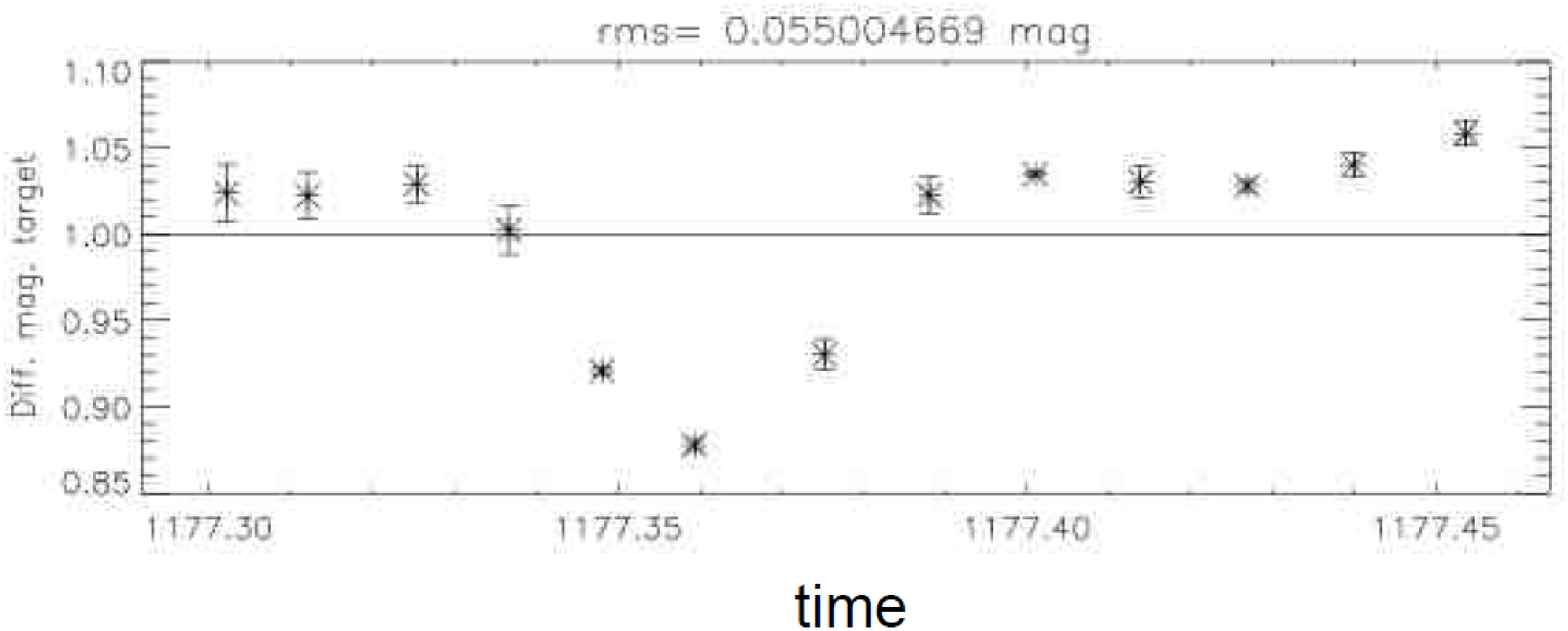} \\
\end{array} $
 \caption{Left: rms photometric precision distribution for a sample of $\sim100$ APACHE targets observed during the first 
 six months of survey operation. 
 Right: APACHE\#1 independent discovery lightcurve of an eclipsing binary first reported by Kiraga (2012).}
\label{fig6}
\end{figure}

Official survey operations started in July 2012, and the project evolution can be followed through its official website 
at http://apacheproject.altervista.org. With six months of data collected in hand, we can begin to look at some 
statistics of the survey. Approximately 100 fields have been observed thus far, with each APACHE telescope typically observing 
a dozen targets each night. The weather pattern of the Summer and Fall seasons has confirmed the expected fraction ($\sim50\%$) 
of useful nights. 

Using the robust TEEPEE pipeline for the reduction and analysis of the
photometric data (Giacobbe et al. 2012), we performed differential aperture photometry on the ensemble of targets 
observed to-date. The resulting distribution of the rms photometric precision for the sample (Fig.~\ref{fig6}, left panel) 
has a median of $\sim 5$ mmag, attesting of the quality of the data. We independently discovered (Fig.~\ref{fig6}, right panel) a new eclipsing binary 
amongst our targets, that was published by Kiraga (2012) while we were performing a preliminary analysis on the data. 

In parallel, we are developing the pipeline TSE (TEEPEE Sentinel) which is designed to perform real-time
differential photometry of the targets during each night. Once fully developed and tested,
TSE will be used for a prompt identification and assessment of interesting signals, triggering real-time changes in the observing strategy 
such as high-cadence follow-up of ongoing possible events (magnifying the sensitivity to long-period planets).

A significant update to the AIC is now ongoing, based on the latest spectroscopic information for many of the APACHE targets 
published by L\'epine et al. (2013). The repercussions are particularly non-negligible in the M0-M1 sub-spectral type regime, 
with many objects being spectroscopically recognized as earlier-type dwarfs. 

\section{Summary}

APACHE is the first Europe-based transit search for small-radius planets orbiting, bright, nearby M dwarfs. It draws on the 
pioneering work carried out by the MEarth project (Nutzman \& Charbonneau 2008), in that is adopts a one target per field approach and the correspoding 
constraints on the observing strategy optimized for the array of five APACHE telescopes and the chosen site characteristics. 
It extends and complements MEarth in the Northern hemisphere, in that it targets early- to mid-M dwarfs, conceding in 
detectable planet radius for fixed photometric precision but focusing  by construction on a target sample that's in principle 
more favourable for atmospheric characterization work in case of positive detection. After the first six months of operations, 
the program is delivering the expected quality in differential photometric precision. The synergy potential between APACHE, 
Doppler (e.g., HARPS-N), and astrometric (e.g., Gaia) techniques is potentially huge (see, e.g., Sozzetti, this volume), 
and the APACHE optical lightcurves of thousands of M dwarf stars will contribute directly to create a database of uniquely well-characterized 
low-mass stars in the solar neighborhood, inclusive of exoplanets with precise mass and radius determinations orbiting M dwarfs bright enough to enable detailed 
characterization of the planetary interior structure and atmospheres with upcoming space-borne observatories (e.g., CHEOPS, EChO, JWST). 

\begin{acknowledgement}
PC, AB, AC, DC and JMC are supported by grants provided by the European Union, the
Autonomous Region of the Aosta Valley and the Italian Department for Work, Health and Pensions.
MD and PG acknowledge support by ASI under contract to INAF I/058/10/0 (Gaia Mission - The Italian Participation to DPAC).
The OAVdA is supported by the Regional Government of the Aosta Valley, the Town Municipality
of Nus and the Mont Emilius Community. We gratefully acknowledge partial support for the APACHE infrastructure 
by the Fondazione Cassa di Risparmio di Torino (CRT). The OAVdA is supported by the Regional Government of the Aosta Valley, 
the Town Municipality of Nus and the Monte Emilius Community.

\end{acknowledgement}

\end{document}